\documentclass[12pt,a4paper]{article}
\usepackage{amsmath,amssymb,amsfonts,amscd,bbm}
\usepackage{graphicx}
\usepackage{cite}
\begin{document} 

%%%%%%%%%%%%%%%%%%%%%%%%%%%%%%%%%%%%%%%%%%%%%%%%%%%%%%%%%%%%%%%%%%%%%
%                     Titel Page                                    %
%%%%%%%%%%%%%%%%%%%%%%%%%%%%%%%%%%%%%%%%%%%%%%%%%%%%%%%%%%%%%%%%%%%%%

\thispagestyle{empty}
\renewcommand{\thefootnote}{\fnsymbol{footnote}}
\setcounter{footnote}{1}

\vspace*{-1.cm}
\begin{flushright}
TUM-HEP-503/03\\
DO-TH 03/05
\end{flushright}
\vspace*{1.8cm}

\centerline{\Large\bf Seesaw Neutrino Masses with Large Mixings}
\vspace*{3mm}
\centerline{\Large\bf from Dimensional Deconstruction}

\vspace*{18mm}

\centerline{\large\bf 
K.R.S. Balaji$^a$\footnote{E-mail: \texttt{balaji@hep.physics.mcgill.ca}}, 
Manfred Lindner$^b$\footnote{E-mail: \texttt{lindner@ph.tum.de}},
and 
Gerhart Seidl$^b$\footnote{E-mail: \texttt{gseidl@ph.tum.de}}}
      
\vspace*{5mm}
\begin{center}
{\em $^a$ Institut f\"ur Theoretische Physik, Universit\"at Dortmund,}\\
{\em Otto-Hahn-Stra{\ss}e 4, 44221 Dortmund, Germany}\\[3mm]
{\em $^b$ Institut f\"ur Theoretische Physik, Physik-Department,\\ 
Technische Universit\"at M\"unchen,}\\
{\em James-Franck-Stra{\ss}e, 85748 Garching, Germany}
\end{center}

\vspace*{20mm}

\centerline{\bf Abstract}
We demonstrate a dynamical origin for the dimension-five seesaw operator 
in dimensional deconstruction models. Light neutrino masses arise  
from the seesaw scale which corresponds to the inverse lattice spacing. 
It is shown that the deconstructing limit naturally prefers maximal 
leptonic mixing. Higher-order corrections which are allowed by gauge 
invariance can transform the bi-maximal into a bi-large mixing.
These terms may appear to be non-renormalizable at scales smaller 
than the deconstruction scale.

%\keywords{Keyword1; keyword2; keyword3.}

\renewcommand{\thefootnote}{\arabic{footnote}}
\setcounter{footnote}{0}

\newpage

%%%%%%%%%%%%%%%%%%%%%%%%%%%%%%%%%%%%%%%%%%%%%%%%%%%%%%%%%%%%%%%%%%%%%
%                     Introduction                                  %
%%%%%%%%%%%%%%%%%%%%%%%%%%%%%%%%%%%%%%%%%%%%%%%%%%%%%%%%%%%%%%%%%%%%%
\section{Introduction}
In contrast to the quark sector, the present state-of-the-art neutrino 
experiments \cite{sksno} demand large or even maximal mixings.
An attempt to unification requires new physics beyond the standard model (SM). 
For instance, in  SO(10) models, one can relate the spectra of both the charged and 
neutral fermions in agreement with known phenomenology \cite{babubarr}. Broadly, most 
efforts to explain flavour physics are either 
(i) strongly model dependent (see e.g. \cite{barr2000}) or (ii) require initial 
assumptions on the neutrino spectra (see e.g. \cite{bala2000}). 
Another option is to explore the framework of dimensional deconstruction
\cite{ark001}, where the effects of higher dimensions originate as a pure dynamical effect in the 
infrared limit. In this context, it is interesting to study the impact on the Yukawa 
sector of a model \cite{seidl003}. Here, we recover the salient aspects of 
neutrino phenomenology, namely large mixings and light masses, from a completely massless 
four dimensional theory at some large scale. The lightness of the 
neutrino masses are an outcome of 
deconstruction which projects out the dimension-five seesaw operator
\cite{yana79}. This scenario contains massless Nambu-Goldstone modes 
corresponding to a symmetry which can be associated with large mixings. All 
of these basic features can be easily understood by considering a simple 
two-site lattice model. 
%%%%%%%%%%%%%%%%%%%%%%%%%%%%%%%%%%%%%%%%%%%%%%%%%%%%%%%%%%%%%%%%%%%%%
%                     Framework                                     %
%%%%%%%%%%%%%%%%%%%%%%%%%%%%%%%%%%%%%%%%%%%%%%%%%%%%%%%%%%%%%%%%%%%%%
\section{The two-site model}
\label{twosite}
\begin{figure}
\begin{center}
\includegraphics*[bb = 228 646 355 717]{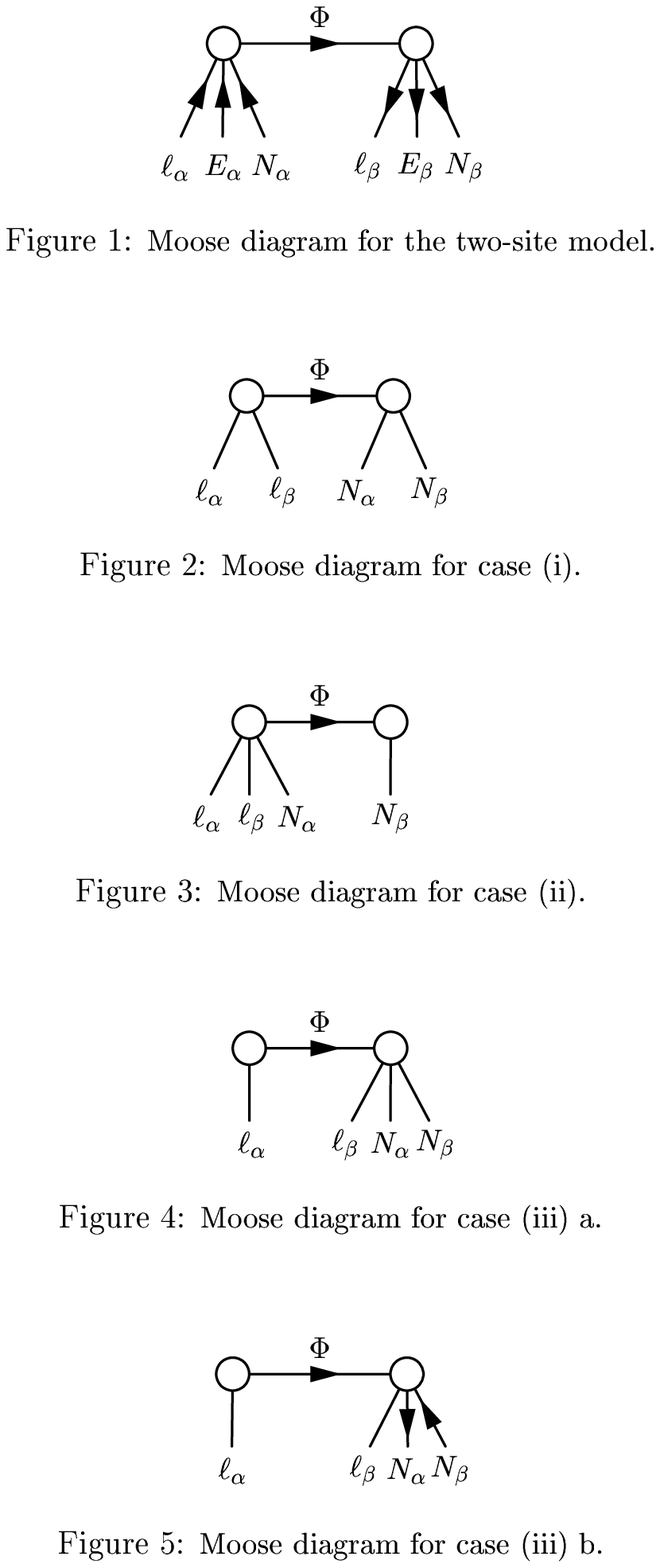}
\end{center}
\vspace*{-8mm}
\caption{\small{The two-site model.}}
  \label{fig:twosites}
\end{figure}

Consider a 
$G = G_{SM}\times SU(m)_1\times SU(m)_2$ gauge theory for deconstructed extra 
dimensions, where $G_{SM}$ denotes the SM 
gauge group. The left-handed lepton doublets are denoted by 
$\ell_\alpha=(\nu_{\alpha L},\:e_{\alpha L})^T$ and the corresponding 
right-handed charged leptons by $E_\alpha$, where the Greek indices denote the usual
flavors $(e,~\mu~\mbox{and}~\tau)$. We will assume that 
$\ell_\alpha$ and $E_\alpha$ 
transform as $\overline{m}_1$ under $SU(m)_1$ and 
$\ell_\beta$ and $E_\beta$ transform as $m_2$ under $SU(m)_2$. We introduce the 
right-handed neutrinos 
$N_{\alpha}$ and $N_{\beta}$, where $N_{\alpha}$ transforms as 
$\overline{m}_1$ under $SU(m)_1$ while $N_{\beta}$ transforms as $m_2$ under 
$SU(m)_2$. The scalar link field $\Phi$ connects as the bi-fundamental
representation $(m_1,\overline{m}_2)$ the neighboring $SU(m)_i$ groups.
This field theory is summarized by the ``moose''
or ``quiver'' \cite{geor86} diagram in Fig.~\ref{fig:twosites}.
The most general renormalizable Yukawa interactions for the neutrinos are then given by
\begin{equation}\label{Yukawa1}
 \mathcal{L}_Y=Y_{\alpha}\overline{\ell_\alpha}\tilde{H}N_{\alpha}
 +Y_{\beta}\overline{\ell_\beta}\tilde{H}N_{\beta}+
f\overline{N_{\alpha}^c}
 \Phi N_{\beta}
 +{\rm h.c.}~.
\end{equation}
 The kinetic term for the link field is 
$\sim (D_\mu \Phi)^\dagger D^\mu \Phi
~;~ D_\mu\Phi= (\partial_\mu -ig_1A_{1\mu}^a T_a+ig_2A_{2\mu}^aT_a)
\Phi~$, where $A_{i\mu}^a$ $(i=1,2)$ are the gauge fields and $T_a$ represent the 
group generators along with the dimensionless gauge couplings, $g_1$ 
and $g_2$. In (\ref{Yukawa1}), $\tilde{H}=i\sigma^2 H^\ast$ is the charge 
conjugated Higgs doublet and $Y_{\alpha},Y_{\beta},f$ are complex 
Yukawa couplings of $\mathcal{O}(1)$. Note that in (\ref{Yukawa1}) the bare Dirac and 
Majorana mass terms of the types $\sim\overline{N_{\alpha}^c}N_{\beta}$ 
and $\sim\overline{N^c_{\alpha}}N_{\alpha}$ or
$\sim\overline{N^c_{\beta}}N_{\beta}$ are forbidden by invariance
under the group $G$. For a suitable scalar potential the field
$\Phi$ can acquire a VEV such that $\langle \Phi\rangle=M_x$, thereby
generating a mass for the scalar field. $M_x$ is identified with 
the deconstruction scale at which the $SU(m)_1\times SU(m)_2$ 
symmetry is broken down to the diagonal $SU(m)$, thereby eating one adjoint
Nambu-Goldstone multiplet in the process. The corresponding 
lattice spacing is $a\sim 1/M_x$. After spontaneous symmetry breaking,
the light effective Majorana mass matrix takes the form
\begin{equation}
\mathcal{M}_\nu=Y_{\alpha} Y_{\beta}\frac{\epsilon^2}{fM_x}
\left(
\begin{matrix}
 0 &1\\
 1 & 0
\end{matrix}
\right)~.
\label{matrix2}
\end{equation}
Here, $\epsilon\equiv\langle H\rangle\simeq 10^2\:{\rm GeV}$ is the
electroweak scale and $M_x\simeq 10^{15}\:{\rm GeV}$ is the seesaw scale. 
This simple analysis leads to our main observations. It is not difficult to 
identify the mass matrix in (\ref{matrix2}) as the
one which arises from the usual dimension-five operator of the type
$\sim \nu\nu HH$. In other words, for length scales $r \gg a \sim 1/M_x$, the
renormalizable and gauge invariant Yukawa interaction in (\ref{Yukawa1}) 
reproduces the effects of the fifth dimension. Whereas, for $r\ll a$, we 
retain a completely renormalizable four-dimensional interaction as 
defined in (\ref{Yukawa1}). In addition, contrary to the conventional seesaw 
operator, dimensional deconstruction can naturally lead to maximal mixings
between the two active neutrino flavors, $\nu_\alpha$ and $\nu_\beta$. This 
is realized due to the $\Phi$ field which mediates a symmetry between each of the 
fermions $(N_{\alpha,\beta})$. This symmetry can be interpreted as an interaction 
which conserves a charge $L_\alpha - L_\beta$ which is reflected in the resulting mass 
matrix for $N_{\alpha,\beta}$. This is retained after symmetry breaking 
as there exists the diagonal subgroup 
$SU(m)$ which respects the symmetry such that the $\Phi$ field would 
transform as $(m,\overline m)$. In the gauge sector, this unbroken 
symmetry corresponds to the presence of a zero mode and is 
$A^{a(0)}_\mu \sim (g_2 A_{1\mu}^a+g_1A^a_{2\mu})$.
The Dirac sector of the model remains diagonal due to the nature of this
construction while maximal mixings 
are introduced from the heavy Majorana sector of the resulting 
seesaw operator. Equivalently, the
qualitative features of this system are not altered even if one allows for
both the fermions and scalars to be link variables. We use this freedom
when we discuss the phenomenology for this mechanism in 
section \ref{threesite}. 

\begin{figure}
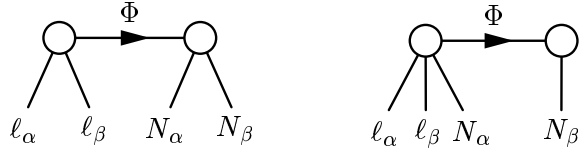

\begin{center}
\includegraphics*[bb = 245 528 343 589]{cases}
\hspace{1cm}
\includegraphics*[bb = 242 411 330 471]{cases}
\end{center}
\vspace*{-8mm}
\caption{\small{Cases (i) (left panel) and (ii) (right
panel).}}
  \label{fig:case1and2}
\end{figure}

\begin{figure}
\begin{center}
\includegraphics*[bb = 258 292 345 354]{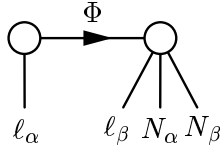}
\end{center}
\vspace*{-8mm}
\caption{\small{Case (iii).}}
  \label{fig:case3}
\end{figure}

Next, we would like to understand how generic is the interaction described in 
(\ref{matrix2}). To answer this question, we consider three possible modifications to
Fig.~\ref{fig:twosites} which are 
summarised in Fig. \ref{fig:case1and2} and Fig. \ref{fig:case3} as 
cases (i)-(iii).
Clearly, in case (i), only higher-order terms of the form 
$\sim \overline{\ell_\alpha} \tilde H \Phi N_{\beta}$ are possible, but a priori, there is
no information on mixings or masses. In case (ii), we have an interaction $\sim 
 Y_{\alpha}\overline{\ell_\alpha}\tilde{H}N_{\alpha}
 +Y_{\beta}\overline{\ell_\beta}\tilde{H}N_{\alpha}+
f\overline{N_{\alpha}^c}
 \Phi N_{\beta}$ which leads to $\mathcal{M}_\nu=0$. Interestingly, for 
case (iii), depending on the representation of the fermionic fields, 
we can envisage two distinct interactions. The first one is of the type $\sim 
 Y_{\alpha}\overline{\ell_\beta}\tilde{H}N_{\alpha}
 +Y_{\beta}\overline{\ell_\beta}\tilde{H}N_{\beta}$
which gives Dirac masses with arbitrary masses and mixings. The second possibility is 
of the type $\sim 
 Y_{\alpha}\overline{\ell_\beta}\tilde{H}N_{\alpha}
 +fM_x \overline {N_{\alpha}^c} N_{\beta} $ 
which again results in $\mathcal{M}_\nu=0$. Note that in the latter
case, gauge invariance allows for a bare mass term which is in contrast to 
(\ref{Yukawa1}).
From the different cases (i)-(iii) we observe a restrictive pattern for the 
allowed fermion masses; this is unlike 
(\ref{Yukawa1}) which ensures a renormalizable 
mass term for all of the resulting Dirac and Majorana fermions. This maximises the 
allowed Yukawa interactions and leads also to maximal mixings.  
In a realistic framework, the basic structure of (\ref{Yukawa1}) is always expected to 
be borne out as we shall demonstrate.
\section{A realistic model}
\label{threesite}
\begin{figure}
\begin{center}
\includegraphics*[bb = 235 529 349 639]{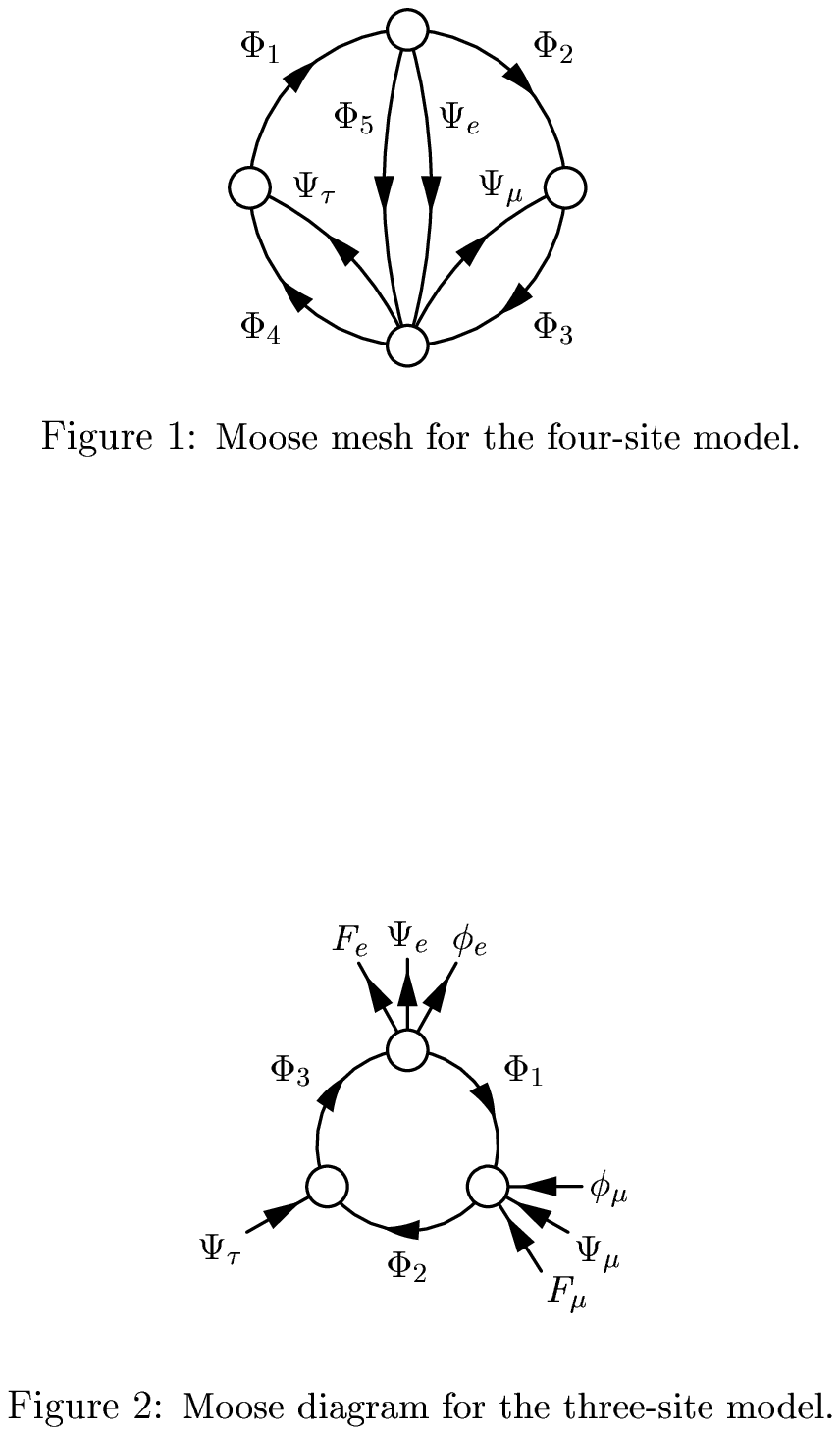}
\end{center}
\vspace*{-8mm}
\caption{\small{A four-site model.}}
  \label{fig:moosemesh}
\end{figure}

We examine a generalization to the case of a moose mesh \cite{greg002}.
Consider a $\Pi_{i=1}^4SU(m)_i$ gauge theory containing five scalar
link variables $\Phi_i$ $(i=1,\ldots,5)$ and the fermion fields $\Psi_\alpha$ is the 
set $\left\{\ell_\alpha,E_\alpha,N_\alpha\right\}$. This model is depicted in 
Fig.~\ref{fig:moosemesh} and up to mass dimension six, we have 
\begin{eqnarray}\label{phenoyuk}
 \mathcal{L}_Y &=& \sum_{\alpha}
  Y_{\alpha}\overline{\ell_\alpha}\tilde{H}N_{\alpha}
 +f_1\overline {N_e^c} \Phi_2^\ast N_\mu+f_2\overline {N_e^c}
 \Phi_1N_\tau+\frac{f_3}{\Lambda}\overline {N_\mu^c} (\Phi_3)^2 N_\mu\nonumber\\
&&+\frac{f_4}{\Lambda}\overline {N_\tau^c} (\Phi_4^\ast)^2 N_\tau+
\frac{f_5}{\Lambda}\overline {N_\mu^c} \Phi_3 \Phi_4^\ast N_\tau+
\frac{f_6}{\Lambda}\overline{N_e^c}(\Phi_5^\ast)^2N_e+\ldots
 +{\rm h.c.},
\end{eqnarray}
where the dots represent non-renormalizable interactions of the leptons
with effective scalar operators involving only the fields $H$ and/or $\Phi_i$. 
In (\ref{phenoyuk}), $Y_{\alpha}$ and $f_i$ $(i=1,\ldots,6)$ are complex couplings 
and $\Lambda (\gg \langle\Phi_i\rangle)$ denotes the scale such that, for 
lattice spacing $a \ll 1/\Lambda$, the theory is fully renormalizable. After symmetry 
breaking and giving universal VEVs
$\langle\Phi_i\rangle\equiv M_x$, the Dirac and Majorana mass 
matrices take the form 
\begin{equation}\label{matrix1}
\mathcal{M}_D=\epsilon
\left(
\begin{matrix}
 Y_{e} &\lambda^2& \lambda^2\\
 \lambda^2&Y_{\mu}&\lambda^2\\
 \lambda^2&\lambda^2 &Y_{\tau}
\end{matrix}
\right),\quad
\mathcal{M}_R=M_x
\left(
\begin{matrix}
 \lambda f_6 & f_1& f_2\\
 f_1&\lambda f_3&\lambda f_5\\
 f_2&\lambda f_5&\lambda f_4
\end{matrix}
\right),\quad\lambda = \frac{M_x}{\Lambda}<1~,
\end{equation}
where only the order of magnitude of the terms with mass dimension $\geq 6$ has
been indicated. We note that in (\ref{matrix1}), as a consequence of the lattice 
geometry $\mathcal{M}_D$ is nearly diagonal while the Majorana sector
carries the $\overline L=L_e-L_\mu-L_\tau$ symmetry which is softly broken by a nonzero
$\lambda$. In the limit 
$\lambda \to 0$, both (\ref{Yukawa1}) and (\ref{phenoyuk}) reproduce similar features.
%\subsection{Numerical estimates}\label{numerics}
Neglecting the small mixing in the Dirac sector\footnote{The contributions
from the charged lepton sector are identical to $M_D$ and can be neglected.} 
and setting $Y_\mu \simeq Y_\tau$ alongwith real couplings, 
$f_1=-f_2=f_3=f_4=f_5\equiv f$, the effective light neutrino mass matrix 
comes to a familiar pattern \cite{grim001} with
\begin{equation}\label{numatrix}
\mathcal{M}_\nu\simeq
\frac{Y_\mu\epsilon^2}{4f^3M_x}\left(
\begin{matrix}
 0 & 2\lambda Y_{e}f^2 & -2\lambda Y_ef^2\\
 2\lambda Y_ef^2 & Y_\mu f(\lambda^2f_6-f) & -Y_\mu f(\lambda^2f_6+f)\\
-2\lambda Y_ef^2 & -Y_\mu f(\lambda^2f_6+f) & Y_\mu f(\lambda^2f_6-f)
\end{matrix}
\right)+ \mathcal{O}(\lambda^3)\:~.
\end{equation}
The relations between the solar and atmospheric mass-squared differences,
$\Delta m_\odot^2$ and $\Delta m_{\rm atm}^2$ respectively, and the solar
mixing angle $\theta_{12}$ are
\begin{eqnarray}\label{data}
\frac{\Delta m_\odot^2}{\Delta m_{\rm atm}^2}&\simeq&
2\sqrt{2}\lambda^3\left(\frac{Y_e}{Y_\mu} \frac{f_6}{f}\right)
+\mathcal{O}(\lambda^4)~,\nonumber\\
{\rm tan}\:\theta_{12}&\simeq&1-\frac{\lambda}{2\sqrt{2}}
\left(\frac{Y_\mu}{Y_e}\frac{f_6}{f}\right)
+\frac{\lambda^2}{16}\left(\frac{Y_\mu}{Y_e}\frac{f_6}{f}\right)^2
+\mathcal{O}(\lambda^4)~.
\end{eqnarray}
For illustration, we choose $\lambda=0.22$, $Y_e=f$, $Y_\mu=f_6$, 
$Y_{\mu}/Y_e = 2.5$ and we 
obtain an atmospheric mixing angle $\theta_{23}\simeq \pi/4$ and a
reactor mixing angle close to zero, {\it i.e.}, $U_{e3}\simeq 0$. Such an
allowed choice minimally alters the basic features of a renormalizable
Lagrangian, leading to a soft 
breaking of the $\bar L$ symmetry. Furthermore,
taking $\Delta m_{\rm atm}^2=2.5\times 10^{-3}\:{\rm eV}^2$, we obtain a normal
neutrino mass hierarchy with
$\Delta m_\odot^2 \simeq 7.5 \times 10^{-5}\:{\rm eV}^2$ 
and $\theta_{12}\simeq 32^\circ$ which is in agreement with the MSW LMA-I 
solution \cite{fogli002}. For the above values, the system predicts an 
effective 
neutrinoless double beta decay mass, $m_{ee} \simeq 10^{-3}\:{\rm eV}$. 
 
%\subsection{Alternative models}
We briefly outline two different variations to (\ref{phenoyuk}).
\begin{figure}
\begin{center}
\includegraphics*[bb = 226 252 361 371]{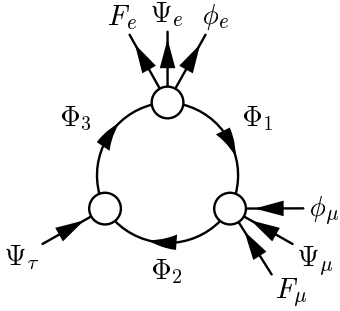}
\end{center}
\vspace*{-8mm}
\caption{\small{A three-site model.}}
  \label{fig:threesites}
\end{figure}
Let us first consider a $\Pi_{i=1}^3 SU(3)_i$ product gauge group with a
representation content as specified in Fig. \ref{fig:threesites} where
the arrows define as before the field transformations. 
%Upon symmetry breaking, the 
%gauge group is broken down to the diagonal $SU(3)$ preserving the conserved 
%lepton number $\overline{L}$. 
We use the Froggatt-Nielsen
mechanism \cite{frog79} to break the $\overline L$ symmetry by putting on 
each of the sites $SU(3)_1$ and $SU(3)_2$
two extra SM singlet fields; these are two scalars $(\phi_e,~\phi_\mu)$ and
two heavy Dirac fermion fields $(F_e~,F_\mu)$.
To retain soft-breaking of the $\overline L$ 
symmetry, we need to impose a $Z_4$ symmetry $\Psi_\alpha\longrightarrow
-\Psi_\alpha$, $\phi_\alpha\longrightarrow -\phi_\alpha$,
$F_{\alpha L }\longrightarrow {\rm i}F_{\alpha L}$,
$\Phi_3\longrightarrow -\Phi_3$, where $\alpha=e,\mu$. We assume that the $SU(3)_1$ 
and $SU(3)_2$ symmetries are broken by bare Majorana mass terms
$\sim\overline{F_{\alpha R}^c}M_{\alpha} F_{\alpha R}$
at some scale $M_\alpha\gg M_x$. When the fields $\phi_\alpha$ aquire
the VEVs $\langle\phi_\alpha\rangle=M_x$ the heavy right-handed
fermions $F_{\alpha R}$ are integrated out leading to the dimension-five terms
$\sim\lambda f_3 M_x \overline{N^c_{\mu}}N_\mu$
and $\sim \lambda f_6 M_x\overline{N^c_e}N_e$, where $\lambda\simeq M_x/M_\alpha$. 
The right-handed neutrino mass matrix is given by
${\mathcal{M}}_R$ in (\ref{matrix1}) with $f_4,f_5\simeq 0$ and hence one 
obtains the relations as in (\ref{data}).
Alternatively, if we perform the identification, $\Phi_2 \to 0$ and 
$SU(3)_i\rightarrow U(1)_i$ and the fields $(\ell_\alpha, N_\alpha)$
$(\alpha=e,\mu,\tau)$ are assigned appropriate $U(1)$ charges, it is not
difficult to derive a model leading to (\ref{matrix1}). 

In conclusion, we argue that upon deconstruction (i) a light neutrino mass is a 
general result and (ii) maximal mixing is inevitable due
to the specific Yukawa interactions in (\ref{Yukawa1}).  
In the limit of a large lattice site model (of size $N\gg 1$), one can draw comparisons 
to the genuine extra-dimensional scenarios (of radius $R$) with the 
identifications 
to the five-dimensional gauge couplings, $g_5(y_i) \to \sqrt{R/N} g_i$, where
$y_i$ denotes the fifth coordinate. In this analysis, we have limited 
ourselves to describing the physics of a periodic lattice where it is
sufficient to examine the periodic interval of any one Brillouin zone.
In general, we predict a small $U_{e3}$ which depends on the 
pattern of the underlying $\overline L$ symmetry breaking.

\section*{Acknowledgments}
K.B. was supported by the ``Bundesministerium f\"ur Bildung, Wissenschaft,
Forschung und Technologie, Bonn'' under contract no. 05HT1PEA9. M.L. and G.S by the
``Sonderforschungsbereich 375 f\"ur Astroteilchenphysik der Deutschen
Forschungsgemeinschaft'' (M.L. and G.S.). K.B. thanks the physics department 
(TUM) for hospitality during completion of this work.

\end{document}